\documentclass[]{pasj01}

\begin{document} 
\Received{2016/09/07}
\Accepted{2017/01/06}

\title{Multi-Color Simultaneous Photometry of the T-Tauri Star Having A Planetary Candidate CVSO~30}

\author{Masahiro \textsc{Onitsuka}\altaffilmark{1,2}}%
\altaffiltext{1}{National Astronomical Observatory of Japan, NINS, 2-21-1 Osawa, Mitaka, Tokyo 181-8588, Japan}
\altaffiltext{2}{SOKENDAI (The Graduate University for Advanced Studies), 2-21-1 Osawa, Mitaka, Tokyo 181-8588, Japan}
\email{masahiro.onitsuka@nao.ac.jp}

\author{Akihiko \textsc{Fukui}\altaffilmark{3}}
\altaffiltext{3}{Okayama Astrophysical Observatory, National Astronomical Observatory of Japan, NINS, Asakuchi, Okayama 719-0232, Japan}

\author{Norio \textsc{Narita},\altaffilmark{1,4,5}}
\altaffiltext{4}{Department of Astronomy, The University of Tokyo, 7-3-1 Hongo, Bunkyo-ku, Tokyo, 113-0033, Japan}
\altaffiltext{5}{Astrobiology Center, NINS, 2-21-1 Osawa, Mitaka, Tokyo, 181-8588, Japan}

\author{Teruyuki \textsc{Hirano}\altaffilmark{6}}
\altaffiltext{6}{Department of Earth and Planetary Sciences, Tokyo Institute of Technology, 2-12-1 Ookayama, Meguro-ku, Tokyo, 152-8551, Japan}

\author{Nobuhiko \textsc{Kusakabe},\altaffilmark{1,5}}

\author{Tsuguru \textsc{Ryu}\altaffilmark{1,2}}

\author{Motohide \textsc{Tamura},\altaffilmark{1,4,5}}


\maketitle

\begin{abstract}
We present three-band simultaneous observations of a weak-line T-Tauri star CVSO~30 (PTFO~8-8695), which is one of the youngest objects having a candidate transiting planet. The data were obtained with the Multicolor Simultaneous Camera for studying Atmospheres of Transiting exoplanets (MuSCAT) on the 188 cm telescope at Okayama Astrophysical Observatory in Japan.
We observed the fading event in the $g^{\prime}_2$-, $r^{\prime}_2$-, and $z_{\rm s,2}$-bands simultaneously.
As a result, we find a significant wavelength dependence of fading depths of about 3.1\%, 1.7\%, 1.0\% for the $g^{\prime}_2$-, $r^{\prime}_2$-, and $z_{\rm s,2}$-bands, respectively.
A cloudless H/He dominant atmosphere of a hot Jupiter cannot explain this large wavelength dependence.
Additionally, we rule out a scenario by the occultation of the gravity-darkened host star.
Thus our result is in favor of the fading origin as circumstellar dust clump or occultation of an accretion hotspot.
\end{abstract}

\section{Introduction}

CVSO~30b (PTFO~8-8695b) is a candidate transiting hot Jupiter around a weak-line T Tauri star in the Orion OB1a/25-Ori region found by \citet{vanEyken2012}. The age of CVSO~30 is estimated as $\sim 3$ Myr \citep{Briceno2005} making this object the youngest candidate planet.
The host star is an M3 type pre-main-sequence star in the Orion OB1a region at
a distance of $323^{+233}_{-96}$ pc. CVSO~30 has the mass of
$0.44 M_{\odot}$ (using the \cite{Baraffe1998} model) or $0.34 M_\odot$ (using the \cite{Siess2000} model),
the radius of $1.39 R_\odot$, 
and the effective temperature of 3470 K \citep{Briceno2005}.
The candidate planet has the orbital period of $0.448413 \pm 0.000040$ days which is comparable to the rotational period of the host star,
the mass of $M_{\rm p} \leq 5.5 \pm 1.4 M_{\rm Jup}$, the radius of $1.91 \pm 0.21 R_{Jup}$,
and the orbital semi-major axis of $0.00838 \pm 0.00072$ AU \citep{vanEyken2012}.

In contrast to other planetary transiting objects, the observed fading light curves of CVSO~30 change with the observational epoch.
\citet{Barnes2013} explained them in the combination of the precession of planetary orbit's ascending node and the stellar gravity darkening effect assuming synchronization of the stellar rotation and the planetary orbital motion.
\citet{Kamiaka2015} added the photometric data and reanalyzed the light curves using the precession and the gravity-darkening model without the co-rotation.

On the other hand, \citet{Ciardi2015} observed a primary transit at $4.5 \micron$ using Spitzer IRAC without any sign of a secondary eclipse.
In addition, their radial velocity measurement showed no evidence of the Rossiter-McLaughlin effect.
\citet{Yu2015} observed 13 events with $i^\prime$-band and other 13 events with $I+z$-band over three years.
They found that the fading period is not constant.
Multiband simultaneous observations were done for five fading events, of which $r^\prime$- and $I+z$-bands observed simultaneously on two other nights, $i^\prime$- and $g^\prime$-bands observed on another two nights, and $I+z$- and $H$-bands observed on one night.
Their multiband observations yielded that the depth of all but one fading events was deeper at short wavelengths; one of the fading events in the $r^\prime$- and $I+z$-bands showed the same depth.
Additionally, they tried to detect the secondary eclipse at infrared and the Rossiter-McLaughlin effect by high-resolution spectroscopy.
However, they were unable to find the expected signature in both observations but found strong variable H$\alpha$ and Ca H \& K lines.
Hence \citet{Yu2015} argued that the transit-like events were unlikely to be caused by a giant planet but rather caused by either starspots near the rotational pole, circumstellar dust clump transit or occultation of an accretion hotspot.
\citet{Raetz2016} observed 33 fading events and found that the period of the fading event was shortened.
They also observed a fading event in the $B$- and the $R$-bands simultaneously, and found that the light curve showed the same depth.
\citet{Johns-Krull2016} observed excess H$\alpha$ emission of CVSO~30.
The excess H$\alpha$ emission was not detected constantly.
However, the observed velocity motion in transit was almost expected.
They argued that this excess came from the planet in the middle of mass loss.

As presented above, the true nature of CVSO~30b is still unclear, and both planet and non-planet scenarios have been argued.
Thus further observational findings are still important to reveal the identity of CVSO~30b.
We focus on that in the previous studies both the presence and absence of the wavelength dependence in the depth of fading events have been reported \citep{Yu2015,Raetz2016}.
We here report our results of independent three-band simultaneous transit photometry for a fading event of CVSO~30. 
The simultaneity is important because CVSO~30 has a large stellar variability and its fading shape is changing with time.
We use the Multicolor Simultaneous Camera for studying Atmospheres of Transiting exoplanets (MuSCAT) \citep{Narita2015} on the 188 cm telescope at Okayama Astrophysical Observatory (OAO) to investigate the wavelength dependence of a fading event. 
We describe the observations and the analysis method in sections~\ref{sec:observation} and \ref{sec:analysis}.
Section~\ref{sec:result} shows the observed light curves, the results of transit fitting and wavelength dependence.
In Section~\ref{sec:discussion}, we discuss the origin of transit-like events.
Finally, we summarize our results and future plans in section~\ref{sec:conclusion}.

\section{Observation}
\label{sec:observation}

We observed a fading event of CVSO~30 on 2016 Feb. 9 (UT). We used the MuSCAT on the 188-cm telescope at OAO.
MuSCAT can simultaneously take three-band images using two dichroic mirrors and the SDSS $g^{\prime}_2$-, $r^{\prime}_2$-, and $z_{\rm s,2}$-band filters. MuSCAT equips three $1024 \times 1024$ pixels CCDs with the field of view of $6^{\prime}.1 \times 6^{\prime}.1$ .

To achieve the high precision photometry, we did not dither but fixed the target position on the detector, and defocused the stellar PSF.
The exposure times were 60 seconds for all the bands.
We obtained 238, 236, and 236 frames for the $g^{\prime}_2$-, $r^{\prime}_2$-, and $z_{\rm s,2}$-bands, respectively, in 4.4 hours.
Data reduction and aperture photometry were processed using the customized pipeline by \citet{Fukui2011}.

\section{Analysis}
\label{sec:analysis}

Our analysis method is similar to that used by \citet{Fukui2016}.
First, we make light curves using different combinations of comparison stars and various aperture radii.
We choose the comparison stars which are not saturated nor variable stars and the aperture radius which produces a light curve with minimum dispersion outside the fading event.
The selected aperture radii are 17, 14, and 12 pixels for the $g^{\prime}_2$-, $r^{\prime}_2$-, and $z_{\rm s,2}$-bands, respectively, and the comparison stars are the same two stars for each band.
Second, we convert the time system of Modified Julian Day (MJD) in Universal Time Coordinate (UTC)
which is recorded in the FITS headers into Barycentric Julian Day (BJD) in Barycentric Dynamical Time (TDB)
using the algorithm of \citet{Eastman2010}. 
We also remove outliers that separate beyond $5 \sigma$ from mean relative flux of each band.
As a result, we use 172, 221, and 220 data for the $g^{\prime}_2$-, $r^{\prime}_2$-, and $z_{\rm s,2}$-bands, respectively.
The produced light curves are shown in the top panels in Figure \ref{fig:lc_obs}.

Next, we fit the light curves with a transit model, assuming that the fading event is caused by a transiting spherical body.
We use the transit light curve model by \citet{Ohta2009} which is comparable to the quadratic limb-darkening law case of \citet{Mandel2002}. 
The transit modeling parameters are the time of the transit center $T_{\rm c}$, the ratio of the planet to the stellar radius $R_{\rm p}/R_{\rm s}$, the ratio of the semimajor axis to the stellar radius $a/R_{\rm s}$, and the impact parameter $b$.
The variations seen outside the fading event seem to be caused by the variability of CVSO~30 itself or comparison stars, with changing of airmass, shifting of stellar images on the detector and so on.
We fit the transit light curves and baseline model at the same time using the customized code by \citet{Fukui2016} which is based on \citet{Narita2007} and \citet{Narita2013}.
A transit and out-of-transit (OOT) model is expressed as
\begin{equation}
 F = ( k_0 + \sum_{i=1} k_i X_i ) \times F_{\rm tr}
\end{equation}
where $F$ is the relative flux, $F_{\rm tr}$ is the transit light curve model, \{{\bf X}\} are time-dependent observed variables for example airmass, shifts of stellar position on the CCD and so on, \{{\bf k}\} are coefficients \citep{Fukui2016}.

We fix the orbital eccentricity to zero and the limb-darkening parameters to the values from \citet{Claret2012}, $\{u1, u2\}=\{0.8067, 0.0861\}, \{0.8423, -0.0160\}, \{0.2946, 0.3569\}$ for the $g^{\prime}_2$-, $r^{\prime}_2$-, and $z_{\rm s,2}$-bands. In addition, we put a Gaussian prior for $a/Rs = 1.685 \pm 0.064$, based on the value of \citet{vanEyken2012}.

For choosing the most appropriate combination of free parameters, we optimize the parameters using the AMOEBA algorithm \citep{Press1992} and evaluate the Bayesian Information Criteria (BIC: \cite{Schwarz1978}).
The BIC value is defined by ${\rm BIC} \equiv \chi^2 + k {\rm ln}N$, where $k$ is the number of free parameters, and $N$ is the number of data points.
We calculate the BIC values for each light curve and parameter, then we pick out the combinations to the minimum BIC value for each band.
For \{{\bf X}\}, we test different combinations of the time $t$, the square of time $t^2$, the airmass $z$, and the relative stellar positions on the CCD $\Delta x, \Delta y$.
Based on minimum BIC, we adopt $k_0$, $k_{\rm t}$ for all bands.

To take into account the time-correlated noise (so-called red noise; e.g. \cite{Pont2006, Winn2008}), we multiply each flux uncertainty by the red-noise factor $\beta$ which is the ratio of the observed standard deviation of the binned light curve to the expected standard deviation of the unbinned and non-time-correlated light curve.
The binning sizes are between 5 and 20 minutes.
We obtain the median of $\beta$ for each binning size and multiply the median $\beta$ by the uncertainty of each data.
The typical photometric errors including $\beta$ are 2.1\%, 0.87\%, 0.54\% for the $g^{\prime}_2$-, $r^{\prime}_2$-, and $z_{\rm s,2}$-bands, respectively. 

After that, to estimate the uncertainty of the free parameters, we analyze the chosen light curve by the Markov Chain Monte Carlo (MCMC) method following \citet{Narita2013} and \citet{Fukui2016}.
We first analyze independently the light curves for each band, and after that we jointly analyze all the light curves.
We make the MCMC chain with $10^7$ steps and first $10^6$ steps are excluded as burn-in.
We define 1$\sigma$ uncertainties as the range of the parameters between 15.87\% and 84.13\% of the merged posterior distributions.

\section{Results}
\label{sec:result}

\begin{figure*}[t]
 \begin{center}
  \includegraphics[width=10cm]{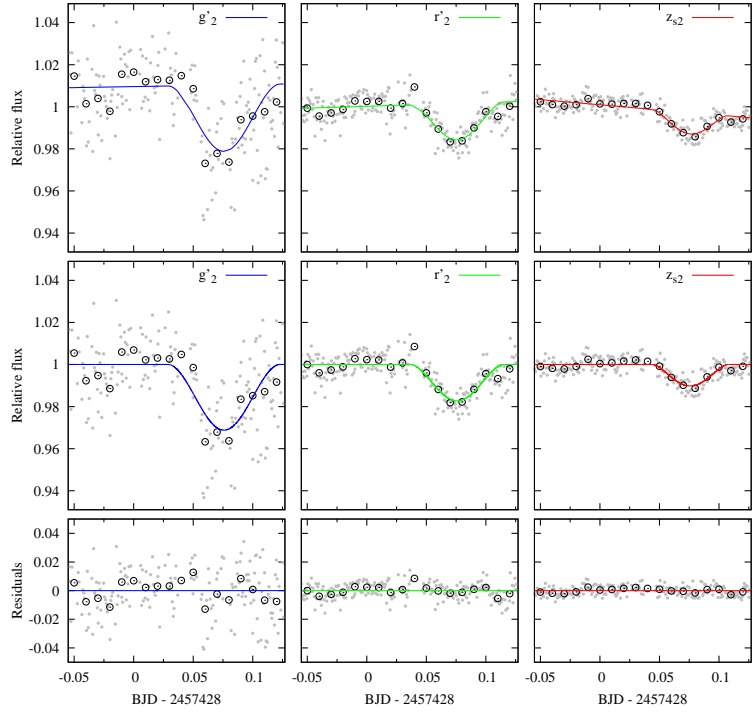} 
 \end{center}
\caption{Top panels: The gray dots are the raw light curves of CVSO~30 observed by $g^{\prime}_2$-, $r^{\prime}_2$-, and $z_{\rm s,2}$-band from the left. The black open circles are 0.01 days binned data. The solid lines are the best-fit transit models. Middle Panels: The light curves corrected by the baseline correction. Bottom panels: The residuals between the observed data and the best-fit model.}\label{fig:lc_obs}
\end{figure*}

Table \ref{tab:param} shows the best-fit transit parameters and uncertainties obtained by the MCMC analysis. 
Note that we obtain consistent results when we fit the light curves that are produced by different baseline models with comparable BIC values.
We show the transit light curves and the best-fit models of each observation in Figure \ref{fig:lc_obs}. 

\begin{table}
  \tbl{The best fit parameters and the uncertainties}{
  \begin{tabular}{lcc}
      \hline
      Parameter & Value & Uncertainty \\ 
      \hline
  $T_{\rm c}$ [BJD$_{\rm TDB}$] & $2457428.0758$ & $^{+0.0017} _{-0.0014}$ \\
  $a/R_{\rm s}$ & $1.730$ & $\pm 0.061$ \\
  $b$ & $1.054$ & $^{+0.121} _{-0.086}$ \\
  $R_{\rm p}/R_{\rm s}$ ($g ^\prime_2$) & $0.344$ & $^{+0.111} _{-0.075}$ \\
  $R_{\rm p}/R_{\rm s}$ ($r ^\prime_2$) & $0.271$ & $^{+0.105} _{-0.066}$ \\
  $R_{\rm p}/R_{\rm s}$ ($z_{\rm s,2}$) & $0.206$ & $^{+0.104} _{-0.064}$ \\
  $k_0$ ($g ^\prime_2$) & $1.0103$ & $^{+0.0055} _{-0.00434}$ \\
  $k_0$ ($r ^\prime_2$) & $1.0016$ & $^{+0.0017} _{-0.0014}$ \\
  $k_0$ ($z_{\rm s,2}$) & $0.99713$ & $^{+0.00080} _{-0.00074}$ \\
  $k_{\rm t}$ ($g ^\prime_2$) & $0.010$ & $^{+0.059} _{-0.049}$ \\
  $k_{\rm t}$ ($r ^\prime_2$) & $0.018$ & $^{+0.019} _{-0.017}$\\
  $k_{\rm t}$ ($z_{\rm s,2}$) & $-0.0489$ & $^{+0.0093} _{-0.0087}$ \\
      \hline
    \end{tabular}}\label{tab:param}
\end{table}

\begin{table}
  \tbl{$R_{\rm p}/R_{\rm s}$ and the uncertainties in case of fixed various impact parameter for each band}{
  \begin{tabular}{lcc}
      \hline
      Filter & Impact parameter $b$ & $R_{\rm p}/R_{\rm s}$ \\ 
      \hline
  $g^\prime_2$ & 0.97 & $0.273 ^{+0.024} _{-0.026}$ \\
  & 1.05 & $0.342 ^{+0.025} _{-0.027}$ \\
  & 1.18 & $0.451 \pm 0.028$ \\
  $r^\prime_2$ & 0.97 & $0.205 ^{+0.011} _{-0.012}$ \\
  & 1.05 & $0.270 \pm 0.012 $ \\
  & 1.18 & $0.376 ^{+0.012} _{-0.013}$ \\
  $z_{\rm s,2}$ & 0.97 & $0.1427 ^{+0.0086} _{-0.0092}$ \\
  & 1.05 & $0.2057 ^{+0.0087} _{-0.0092}$ \\
  & 1.18 & $0.3099 ^{+0.0086} _{-0.0090}$ \\
      \hline
    \end{tabular}}\label{tab:impact}
\end{table}

\begin{figure}
 \begin{center}
  \includegraphics[width=8cm]{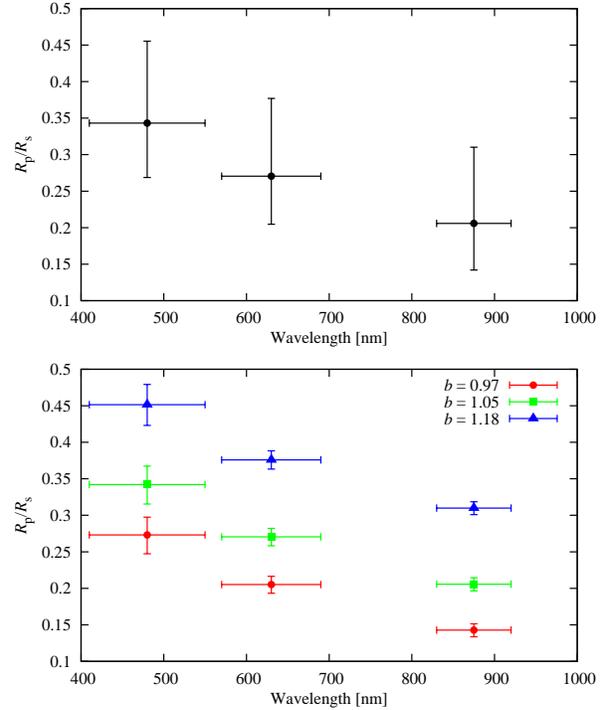} 
 \end{center}
\caption{Top panel: The best-fit parameter and uncertainty of $R_{\rm p}/R_{\rm s}$ in $g^{\prime}_2$-, $r^{\prime}_2$-, and $z_{\rm s,2}$-band from Table \ref{tab:param}. Bottom panel: The wavelength dependence of $R_{\rm p}/R_{\rm s}$ at the impact parameter fixed by $b =\{0.97, 1.05, 1.18\}$}
\label{fig:rprs_comp}
\end{figure}

\citet{vanEyken2012} and \citet{Yu2015} reported that the fading depth of CVSO~30 changes with the observational epoch.
According to \citet{Yu2015}, the loss of light was 20-30\% larger in the bluer band on 2012 Dec. 14 at $r^\prime$- and $I + z$-band, on 2014 Jan. 9 and 2014 Jan. 18 at $i^\prime$- and $g^\prime$-band, and on 2014 Jan. 19 at $I + z$- and $H$-band respectively.
However, the loss of light in $r^\prime$ was essentially the same as in $I + z$ on 2012 Dec. 15.
We present the fading depth in $g^{\prime}_2$-, $r^{\prime}_2$-, and $z_{\rm s,2}$-bands observed simultaneously.
Figure \ref{fig:rprs_comp} gives the wavelength dependence of $R_{\rm p}/R_{\rm s}$.
The top panel shows the best-fit parameters for $R_{\rm p}/R_{\rm s}$ and their uncertainties for each band in Table \ref{tab:param}.
The uncertainties are strongly influenced by the impact parameter $b$ which is grazing.
We recalculate the $R_{\rm p}/R_{\rm s}$ and their uncertainties fixing the impact parameter at the median and 1 $\sigma$ upper/lower limit $b = \{0.97, 1.05, 1.18\}$ in order to highlight the wavelength dependence of $R_{\rm p}/R_{\rm s}$ (Figure \ref{fig:rprs_comp}: bottom panel; Table \ref{tab:impact}).
As the result, we find that the clear wavelength dependence in $R_{\rm p}/R_{\rm s}$ is apparent regardless of the value of $b$.

\section{Discussion}
\label{sec:discussion}
Each of CVSO~30's fading light curve shapes varies in the observational epoch, with some showing large fading and others showing small fading \citep{vanEyken2012}.
Therefore, multicolor simultaneous observations are necessary to discuss wavelength dependence of light curves.
The previous multicolor observations are performed in two bands \citep{Yu2015, Raetz2016}.
In order to reveal precise wavelength dependence, we observe in three bands simultaneously.
The apparent $R_{\rm p}/R_{\rm s}$ is larger at shorter wavelengths, 40-90\% larger for $g^{\prime}_2 - z_{\rm s,2}$, and 20-50\% larger for $r^{\prime}_2 - z_{\rm s,2}$.
Now, we consider whether the cloudless atmosphere can explain this wavelength dependence.
The atmospheric scaleheight $H$ is written as
\begin{equation}
 H = \frac{k_BT}{\mu g}
\end{equation}
where $k_B$ is the Boltzmann constant, $T$ is the atmospheric temperature, $\mu$ is the mean molecular weight which is $\mu \approx 2.3$ for H/He-dominated atmosphere (e.g. \cite{deWit2013}), and $g$ is the local gravity.
We assume that the planet candidate equilibrium temperature of 1800 K, the planetary mass of 0.9 $M_{Jup}$ from upper limit of \citet{Ciardi2015}, the planetary radius of 2.8 $R_{Jup}$ from $R_{\rm p}/R_{\rm s}$ of this work and stellar radius of 1.39 $R_\odot$ \citep{Briceno2005}.
Then, the calculated scale height $H$ is 2300 km or 0.0023 $R_{\rm p}$.
Therefore, the variation between the $r^\prime_2$- and the $z_{\rm s,2}$-band corresponds to 30$H$ and that between the $g^\prime_2$- and the $z_{\rm s,2}$-band is 60$H$.
In comparison, the variation of HD~189733b from \authorcite{Sing2011} (\yearcite{Sing2011}, Fig. 14), the radius variation of HD~189733b between 630 nm and 880 nm corresponding to the $r^\prime_2$- and the $z_{\rm s,2}$-band is 1.5 $H$, between 490 nm and 880 nm corresponding to the $g^\prime_2$- and the $z_{\rm s,2}$-band is 2.5 $H$.
In addition, we compare the wavelength dependence with cloudless Rayleigh scattering following \citet{Southworth2015}.
The slope of the planetary radius depending on wavelength as a function is expressed as
\begin{equation}
 \alpha H= \frac{{\rm d}R_{\rm p}(\lambda)}{{\rm d}\ln \lambda}
\end{equation}
where $\alpha$ is a power-low coefficient ($\alpha = -4$ for Rayleigh scattering), and $\lambda$ is wavelength.
The corresponding differential radius is 63000 km for the $r^\prime_2$-band vs the $z_{\rm s,2}$-band and 130000 km for the $g^\prime_2$-band vs the $z_{\rm s,2}$-band.
This result yields that $\alpha$ is -80 to -100.
Therefore, the wavelength dependence of the CVSO~30's fading event is too large to be explained by the atmospheric Rayleigh scattering of a gas giant planet.

We also discuss the influence of gravity darkening \citep{vonZeipel1924}.
\citet{Barnes2013} and \citet{Howarth2016} explained the changes of the shape of transit light curve with gravity darkening.
The intensity distribution on the gravity-darkened star varies with observing wavelengths, and therefore the transit light curves have wavelength dependence \citep{Barnes2009}.
Now, we test whether the wavelength dependence is explainable in terms of gravity-darkening effect.
We calculate the $g^{\prime}_2$-, $r^{\prime}_2$-, and $z_{\rm s,2}$-band model light curve with gravity darkening in the condition that the largest wavelength dependence is expected.
The derivation of the intensity distribution is obtained by numerical integration with the formula by \citet{Barnes2009}.
We assume the strongest gravity darkening case, the stellar obliquity $90^\circ$ pole-on orbit, the impact parameter $0$, $R_{\rm p}/R_{\rm s} = 0.11$ to approximate the observed fading depth in the $r^{\prime}_2$-band.
We also fix $a/R_{\rm s}$ with the best-fit parameter in Table \ref{tab:param}, limb darkening parameter with the same value in Section \ref{sec:analysis}, the stellar rotational velocity with $120$ km/s obtained by the stellar radius and the rotational period in \citet{vanEyken2012}.

\begin{figure}
 \begin{center}
  \includegraphics[width=8cm]{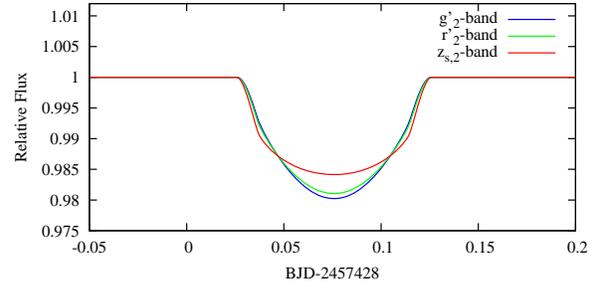} 
 \end{center}
\caption{The model light curve included gravity-darkening effect.}
\label{fig:gravdark}
\end{figure}

The derived gravity-darkened model light curve is shown in Figure \ref{fig:gravdark}.
The transit depths with gravity darkening are 2.0\%, 1.9\%, 1.6\% for the $g^{\prime}_2$-, $r^{\prime}_2$-, and $z_{\rm s,2}$-bands, respectively (Table \ref{tab:gravdark}).
The wavelength dependence caused by gravity darkening is weak and not able to reproduce the observed depths of 3.1\%, 1.7\%, 1.0\% for the $g^{\prime}_2$-, $r^{\prime}_2$-, and $z_{\rm s,2}$-band, respectively.
In other words, the observed wavelength dependence of the fading depths is too large and cannot be explained by the gravity-darkening effect alone.

\begin{table}
  \tbl{Comparison between the observed depth and the gravity-darkened transit model depth}{
  \begin{tabular}{lcc}
      \hline
      Filter & observed depth [\%] & gravity-darkened model depth [\%]\\ 
      \hline
  $g^\prime_2$ & $3.0811^{+0.0151}_{-0.0035}$ & 1.98 \\
  $r^\prime_2$ & $1.7270^{+0.0108}_{-0.0025}$ & 1.89 \\
  $z_{\rm s,2}$ & $1.0191^{+0.0081}_{-0.0019}$ & 1.59 \\
      \hline
    \end{tabular}}\label{tab:gravdark}
\end{table}

The candidate origins of the fading events except a gas giant proposed by \citet{Yu2015} are starspots near the rotational pole, and orbiting or accreting dust.
According to \citet{Grankin2008}, weak-line T-Tauri stars have stable long-term periodic variability for several years by starspots and stellar rotation.
Such long-term stability is, however, the result of starspot modulation on so-called active longitudes despite shorter lifetime (weeks) of each starspot.
CVSO~30 has been observed to display transit-like fading events over six years since the first observations by \citet{vanEyken2012}.
The starspots as a cause of fading events are unlikely because it is difficult for the spots to exist near the pole at all times.

The remaining possibility for the wavelength dependence would be in favor of a transiting dust clump.
As an example of dust with a rocky planet, \citet{Rappaport2012} found that the shape of transit light curves of KIC~12557548 changes in time.
In another example of a disintegrating planet, \citet{Sanchis-Ojeda2015} signified the wavelength difference of the transit depth by spectro-photometric observation in K2-22b.
The exponent of the our extinction power law which is defined as $- {\rm d\,ln} A/{\rm d\,ln} \lambda$, where A is fading depth obtained by $(R_{\rm p}/R_{\rm s})^2$, is $1.7 \pm 0.8$.
This value is similar to the exponent of the extinction by interstellar medium $2.13 \pm 0.08$ obtained by \citet{Damineli2016}.
The candidate sources of dust include a disintegrating rocky planet, a circumstellar disk, and an occultation of accretion hotspot.

\section{Conclusion}
\label{sec:conclusion}

We have observed the transit-like fading event of the weak-line T-Tauri star CVSO~30 in the $g^{\prime}_2$-, $r^{\prime}_2$-, and $z_{\rm s,2}$-bands simultaneously for the first time, using the MuSCAT instrument on the 188 cm telescope at Okayama Astrophysical Observatory.
We perform the light curves fitting using transit model with the independent $R_{\rm p}/R_{\rm s}$, the same $a/R_{\rm s}$, and the same impact parameter for each band.
We have detected significant wavelength dependence in CVSO~30's fading light curve.
The result of transit light curve fitting shows the large wavelength dependence in transit depths of 3.1\%, 1.8\%, 1.1\% for the $g^{\prime}_2$-, $r^{\prime}_2$-, and $z_{\rm s,2}$-bands, respectively.
The wavelength dependence rules out a transiting gas giant scenario because of being too large to be due to a hydrogen-dominated atmosphere of a hot Jupiter or the gravity-darkening effect.
In addition, the starspots as a cause of fading events are unlikely because it is difficult for the spots to exist several years near the pole at all times.
Thus our result is in favor of transit by circumstellar dust clump or occultation of an accretion hotspot which were introduced by \citet{Yu2015}.

\begin{ack}
The authors thank Yasunori Hori for fruitful discussions.
A.F. and T.H. is supported by the Astrobiology Center Project of National Institutes of Natural Sciences(NINS) Grant Numbers AB281012 and JY280092.
N.N. acknowledges a support by Grant-in-Aid for Scientific Research (A) (JSPS KAKENHI Grant Number 25247026).
This work was supported by JSPS KAKENHI Grant Numbers 16K13791 and 16K17660.
M.T. is partly supported by JSPS KAKENHI Grant Number 15H02063.
\end{ack}

\end{document}